\documentclass[a4paper,10pt,twoside]{cpc-hepnp}
\usepackage{CJK,upgreek,fancyhdr}
\usepackage{multicol}
\usepackage{graphicx}
\usepackage{booktabs}
\usepackage{amssymb,bm,mathrsfs,bbm,amscd}
\usepackage[tbtags]{amsmath}
\usepackage{lastpage}
\usepackage{color}

\begin{document}
\begin{CJK*}{GBK}{song}



\title{Theoretical investigation of the antimagnetic rotation in $^{104}$Pd
       \thanks{Supported by National Natural Science
               Foundation of China (11875027, 11505058, 11775112, 11775026, 11775099),
               Fundamental Research Funds for the Central Universities (2018MS058)
               and the program of China Scholarships Council (No. 201850735020)}}

\author{%
      Zhen-Hua Zhang $^{1)}$\email{zhzhang@ncepu.edu.cn}%
}
\maketitle

\address{$^{1}$Mathematics and Physics Department,
               North China Electric Power University, Beijing 102206, China}
\address{$^{2}$Department of Physics and Astronomy,
               Mississippi State University, Mississippi 39762, USA}

\begin{abstract}
The particle-number-conserving method based on the cranked shell model
is used to investigate the antimagnetic rotation band in $^{104}$Pd.
The experimental moments of inertia and reduced $B(E2)$
transition probabilities are reproduced well.
The $J^{(2)}/B(E2)$ ratios are also discussed.
The occupation probability of each orbital close to the Fermi surface
and the contribution of each major shell to the total angular momentum
alignment with rotational frequency are analyzed.
The backbending mechanism of the ground state band in $^{104}$Pd is understood clearly
and the configuration of the antimagnetic rotation after backbending is clarified.
In addition, the crossing of a four quasiparticle states
with this antimagnetic rotation band is also predicted.
By examining the the closing of the four proton hole angular momenta towards the neutron
angular momenta, the ``two-shears-like'' mechanism for this antimagnetic rotation is
investigated and two stages of antimagnetic rotation in $^{104}$Pd are seen clearly.
\end{abstract}

\begin{keyword}
particle-number conserving method, \
pairing correlations, \
antimagnetic rotation
\end{keyword}

\begin{pacs}
21.60.-n, \ 21.60.Cs, \ 23.20.Lv, \ 27.60.+j
\end{pacs}


\maketitle

\begin{multicols}{2}

\section{Introduction}{\label{Sec:Introduction}}

Antimagnetic rotation (AMR)~\cite{Frauendorf1996_Proceedings,Frauendorf2001_RMP73-463},
which is predicted by Frauendorf in analogy to the antiferromagnetism
in condensed matter physics, is an interesting exotic phenomena observed
in some near spherical or weakly deformed nuclei.
In AMR bands, higher angular momentum and energy are obtained by
the ``two-shears-like'' mechanism,
i.e., by simultaneously aligning of the angular momentum vector of two valence
proton (neutron) holes toward that of the valence neutron (proton) particle.
The AMR bands have regular sequences of energy levels differing in spin
by $2\hbar$, which is the same as normal rotation.
However, due to the weakly deformed or nearly spherical core,
they are connected by weak $E2$ transitions.
Moreover, the $B(E2)$ values decrease with increasing spin in AMR bands.

AMR has been investigated both experimentally and theoretically
since it was proposed~\cite{Frauendorf2001_RMP73-463}.
Up to now, evidence of AMR has been observed experimentally
mainly in Cd ($Z=48$) and Pd ($Z=46$) isotopes
including $^{105}$Cd~\cite{Choudhury2010_PRC82-061308R},
$^{106}$Cd~\cite{Simons2003_PRL91-162501},
$^{107}$Cd~\cite{Choudhury2013_PRC87-034304},
$^{108}$Cd~\cite{Simons2005_PRC72-024318,Datta2005_PRC71-041305R},
$^{110}$Cd~\cite{Roy2011_PLB694-322},
$^{101}$Pd~\cite{Sugawara2012_PRC86-034326,Sugawara2015_PRC92-024309,Singh2017_JPG44-075105},
and $^{104}$Pd~\cite{Rather2014_PRC89-061303R}.
Recently, possible AMR bands have also been reported
in Eu ($Z=63$) isotopes including $^{143}$Eu~\cite{Rajbanshi2015_PLB748-387}
and $^{142}$Eu~\cite{Ali2017_PRC96-021304R}.
The existence of AMR still needs
further investigation in
$^{109}$Cd~\cite{Chiara2000_PRC61-034318},
$^{100}$Pd~\cite{Zhu2001_PRC64-041302R},
$^{144}$Dy~\cite{Sugawara2009_PRC79-064321},
and $^{112}$In~\cite{Li2012_PRC86-057305}
by lifetime measurements.
AMR has been investigated theoretically mainly by the
semi-classical particle rotor model~\cite{Clark2000_ARNPS50-1},
and the tilted axis cranking (TAC)
model~\cite{Frauendorf2000_NPA677-115,Peng2008_PRC78-024313,Zhao2011_PLB699-181}.
Especially, many investigations have been performed
within the framework of microscopic-macroscopic
model~\cite{Zhu2001_PRC64-041302R,Simons2003_PRL91-162501,Simons2005_PRC72-024318},
pairing plus quadrupole model~\cite{Chiara2000_PRC61-034318,Frauendorf2001_RMP73-463},
and the covariant density functional theory
(CDFT)~\cite{Zhao2011_PRL107-122501,Zhao2012_PRC85-054310,
Peng2015_PRC91-044329,Jia2018_PRC97-024335} based on the TAC model
(for reviews please see Refs.~\cite{Meng2013_FP8-55, Zhao2018_JEMPE27-1830007}).

In Ref.~\cite{Rather2014_PRC89-061303R}, the ground state band (gsb)
of $^{104}$Pd after backbending is assigned as AMR.
Using the semiclassical particle rotor model, its configuration was assumed to be
$\pi g_{9/2}^{-2} \otimes \nu[h_{11/2}^2,(g_{7/2},d_{5/2})^2]$~\cite{Rather2014_PRC89-061303R}.
However, this band was assigned a different configuration as
$\pi g_{9/2}^{-4} \otimes \nu[h_{11/2}^2,(g_{7/2},d_{5/2})^6]$
in a recent investigation using TAC-CDFT~\cite{Jia2018_PRC97-024335}
with the point coupling effective interaction PC-PK1~\cite{Zhao2010_PRC82-054319}.
Note that these two configurations are all written with respect to the $^{100}$Sn core.
It can be seen that these two configuration assignments have different number of proton $g_{9/2}$ holes,
which are very important to the formation of AMR.
Therefore, it is necessary to clarify the AMR configuration in $^{104}$Pd.
In the present work, the particle-number-conserving (PNC)
method~\cite{Zeng1983_NPA405-1,Zeng1994_PRC50-1388} based on the
cranked shell model (CSM) will be used to investigate the AMR in $^{104}$Pd.
Note that the PNC-CSM has already provided successful descriptions for the AMR bands
in $^{105, 106}$Cd~\cite{Zhang2013_PRC87-054314} and $^{101}$Pd~\cite{Zhang2016_PRC94-034305},
and have shown the important role of pairing correlations on the
moment of inertia (MOI) and the ``two-shears-like'' mechanism.

Different from the traditional Bardeen-Cooper-Schrieffer or Hartree-Fock-Bogoliubov approaches,
in the PNC method, the pairing interaction is diagonalized directly
in a sufficiently large Fock-space~\cite{Wu1989_PRC39-666}.
Therefore, it is a shell-model like approach and the particle-number
is totally conserved from beginning to the end
and the Pauli blocking effects are treated exactly.
The PNC scheme has also been transplanted in relativistic
and non-relativistic mean field models~\cite{Meng2006_FPC1-38,Shi2018_PRC97-034317,
Pillet2002_NPA697-141,Liang2015_PRC92-064325},
and the total-Routhian-surface method~\cite{Fu2013_PRC87-044319}.

This paper is organized as follows.
The theoretical framework of PNC-CSM is presented in Sec.~2.
The results and discussion of the AMR band in $^{104}$Pd are given in Sec.~3.
Finally I summarize this work in Sec.~4.

\section{Theoretical framework}{\label{Sec:PNC-CSM}}

For an axially deformed nucleus, the cranked shell model Hamiltonian reads
\begin{eqnarray}
 H_\mathrm{CSM}
 & = &
 H_0 + H_\mathrm{P}
 = H_{\rm Nil}-\omega J_x + H_\mathrm{P}
 \ ,
 \label{eq:H_CSM}
\end{eqnarray}
where $H_{\rm Nil}$ is the Nilsson Hamiltonian~\cite{Nilsson1969_NPA131-1},
$-\omega J_x$ is the Coriolis interaction,
and $H_{\rm P}$ is the monopole pairing interaction
with effective pairing strength $G$.
$H_{\rm P}$ reads
\begin{eqnarray}
 H_{\rm P}
 & = &
  -G \sum_{\xi\eta} a^\dag_{\xi} a^\dag_{\bar{\xi}}
                        a_{\bar{\eta}} a_{\eta}
  \ ,
\end{eqnarray}
where $\bar{\xi}$ ($\bar{\eta}$) denotes the time-reversal state of $\xi$ ($\eta$).

When treating the pairing correlations, a cranked many-particle configuration
(CMPC) truncation is adopted, which can make sure that the
PNC calculations are both workable and sufficiently
accurate~\cite{Wu1989_PRC39-666,Molique1997_PRC56-1795}.
For the investigation of heavy nuclei,
a dimension of 1000 for both protons and neutrons is enough.
By diagonalizing the $H_\mathrm{CSM}$ in a sufficiently
large CMPC space, sufficiently accurate solutions for low-lying excited eigenstates of
$H_\mathrm{CSM}$ can be obtained, which can be written as
\begin{equation}
 |\Psi\rangle = \sum_{i} C_i \left| i \right\rangle
 \ ,
 \qquad (C_i \; \textrm{is real}), \label{eq:psi}
\end{equation}
where $| i \rangle$ is a CMPC (the eigenstate of $H_0$)
and $C_i$ is the corresponding expanding coefficient.
The expectation value of any one-body operator (e.g.,
angular momentum $J_x$ and quadrupole moment $Q_{20}$)
$\mathcal {O} = \sum_{k=1}^N \mathscr{O}(k)$ can be written as
\begin{equation}
 \left\langle \Psi | \mathcal {O} | \Psi \right\rangle
 =\sum_i C_i^2 \left\langle i | \mathcal {O} | i \right\rangle
 +2\sum_{i<j} C_i C_j \left\langle i | \mathcal {O} | j \right\rangle \ .
\end{equation}
Since $\mathcal {O}$ is a one-body operator,
the matrix element $\langle i | \mathcal {O} | j \rangle$ between two different
CMPCs $|i\rangle$ and $|j\rangle$ is nonzero only when these two CMPCs
differ by one particle occupation~\cite{Zeng1994_PRC50-1388}.
After certain permutations of creation operators, these two CMPCs can be written as
\begin{equation}
 | i \rangle = (-1)^{M_{i\mu}} | \mu \cdots \rangle \ , \qquad
 | j \rangle = (-1)^{M_{j\nu}} | \nu \cdots \rangle \ ,
\end{equation}
where $\mu$ and $\nu$ are two different single-particle states,
and $(-1)^{M_{i\mu}}=\pm1$, $(-1)^{M_{j\nu}}=\pm1$ according to
whether the number of permutation is even or odd.
Therefore, the expectation value of the one-body operator $\mathcal {O}$
can be separated into the diagonal $\sum_{\mu} \mathscr{O}(\mu)$
and the off-diagonal $2\sum_{\mu<\nu} \mathscr{O}(\mu\nu)$ parts
\begin{eqnarray}
\left\langle \Psi | \mathcal {O} | \Psi \right\rangle &=&
  \left( \sum_{\mu} \mathscr{O}(\mu) + 2\sum_{\mu<\nu} \mathscr{O}(\mu\nu) \right) \ , \label{eq:j1}\\
 \mathscr{O}(\mu)
 &=& \langle \mu | \mathscr{O} | \mu \rangle n_{\mu}  \ , \label{eq:j1d} \\
 \mathscr{O}(\mu\nu)
 &=&\langle \mu | \mathscr{O} | \nu \rangle
  \sum_{i<j} (-1)^{M_{i\mu}+M_{j\nu}} C_{i} C_{j} \ ,
  \label{eq:j1od}
\end{eqnarray}
where $n_{\mu}=\sum_{i}|C_{i}|^{2}P_{i\mu}$
is the occupation probability of the single-particle state $|\mu\rangle$
and $P_{i\mu}=0$ (1) if $|\mu\rangle$ is empty (occupied) in $|i\rangle$.

The kinematic MOI $J^{(1)}$ and dynamic MOI $J^{(2)}$ are given by
\begin{eqnarray}
 J^{(1)} = \frac{1}{\omega} \left\langle \Psi | J_x | \Psi \right\rangle \ , \quad
 J^{(2)} = \frac{{\rm d} }{{\rm d} \omega} \left\langle \Psi | J_x | \Psi \right\rangle \ .
\end{eqnarray}

The reduced $B(E2)$ transition probability can be obtained from
the semi-classical approximation as
\begin{equation}
B(E2) = \frac{3}{8}
{\left\langle \Psi | Q_{20}^{\rm p} | \Psi \right\rangle}^2 \ ,
\end{equation}
where $Q_{20}^{\rm p}$ is the proton quadrupole moment and
\begin{equation}
Q_{20} = \sqrt{\frac{5}{16\pi}} (3z^2-r^2) = r^2 Y_{20} \ .
\end{equation}

\section{Results and discussion}{\label{Sec:Results}}

In the present work, Nilsson parameters ($\kappa$ and $\mu$)
for $^{104}$Pd are taken from Ref.~\cite{Bengtsson1985_NPA436-14}.
The deformation parameter $\varepsilon_2 = 0.18$ is taken
as the experimental value~\cite{Rather2014_PRC89-061303R}.
The valence single-particle space is constructed
from $N=0$ to $N=5$ major shells both for protons and neutrons.
Note that $N = 0$ to $N = 3$ major shells are closed shells ($N=Z=40$),
and are fully occupied, so their contribution to
the total angular momentum alignment is zero both for protons and neutrons.
However, they are important for calculating the $B(E2)$ values.
With $N = 0$ to $N = 3$ major shells being considered,
there is no effective charge involved when calculating the $B(E2)$ values.
The effective monopole pairing strengths are determined by the experimental
odd-even binding energy differences and the MOIs, and are connected with
the dimension of the truncated CMPC space, which are about
0.9$\hbar\omega_0$ for protons and 0.8$\hbar\omega_0$ for neutrons, respectively.
In the present calculations, the dimensions of the CMPC space
are chosen as 1000 both for protons and neutrons, in which the CMPCs
with weights larger than $0.1\%$ in the many-body
wave-function (c.f., Eq.~\ref{eq:psi}) are all included.
The effective pairing strengths adopted in this work are
$G_{\rm p}$ = 0.5~MeV for protons and $G_{\rm n}$ = 0.8~MeV for neutrons.

\begin{center}
\centering
\includegraphics[width=0.95\columnwidth]{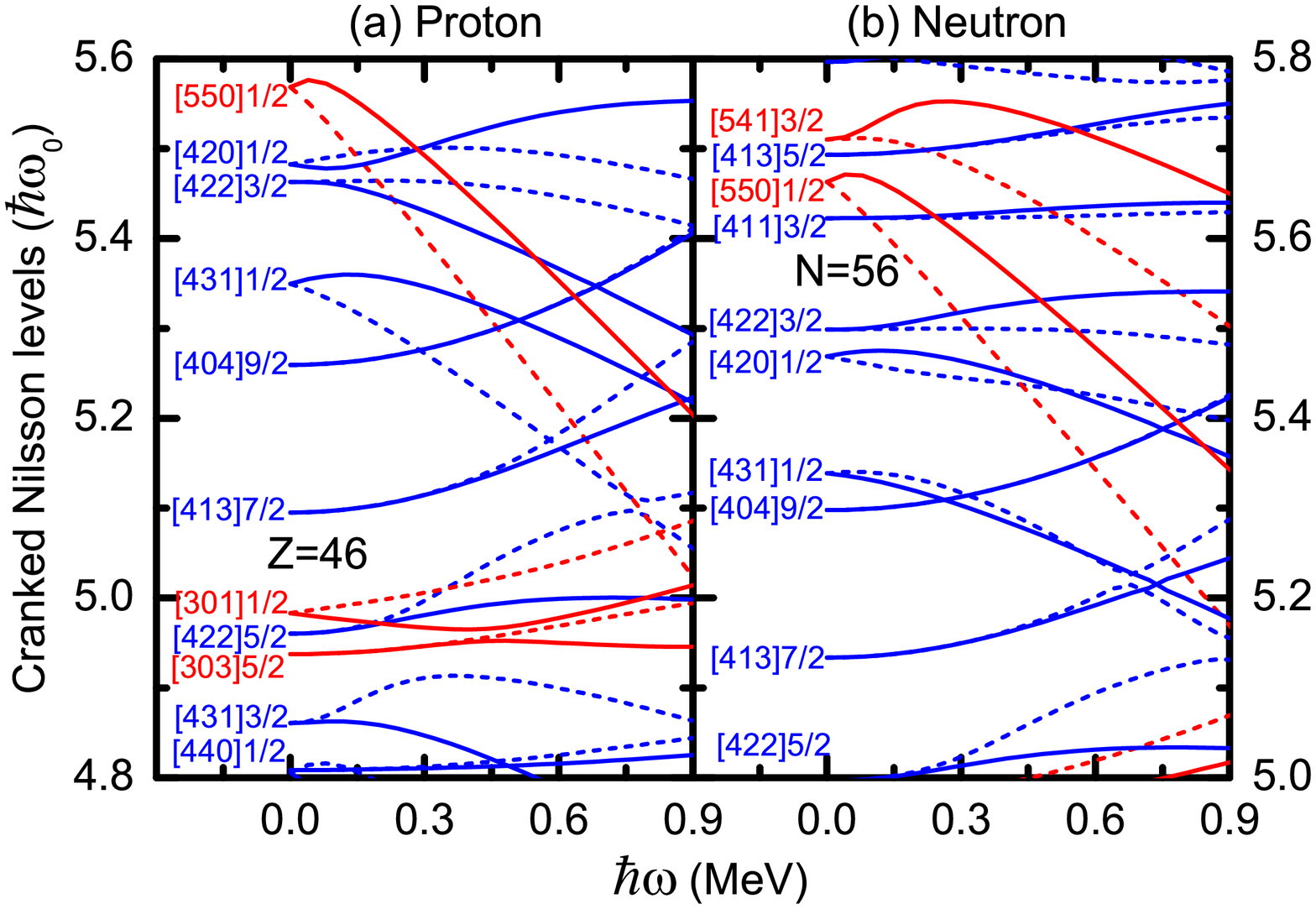}
\figcaption{\label{fig1:Nil}(Color online)
The proton (a) and neutron (b) cranked Nilsson levels near the Fermi surface of $^{104}$Pd.
The positive and negative parity levels are denoted by blue and red lines, respectively.
The signature $\alpha=\pm1/2$ levels
are denoted by solid and dotted lines, respectively.}
\end{center}

The cranked Nilsson levels near the Fermi surface of $^{104}$Pd are shown
in Fig.~\ref{fig1:Nil} for (a) protons and (b) neutrons.
Due to deformation effects, the traditional magic shell gaps with
proton number $Z=50$ and neutron number $N=50$ disappear.
It should be noted that since the quadrupole deformation adopted
in the present calculation for $^{104}$Pd ($\varepsilon_2 = 0.18$)
is much larger than those for $^{105}$Cd ($\varepsilon_2 = 0.12$),
$^{106}$Cd ($\varepsilon_2 = 0.14$), and $^{101}$Pd ($\varepsilon_2 = 0.125$)
in our previous works~\cite{Zhang2013_PRC87-054314, Zhang2016_PRC94-034305},
their single particle structures close to the Fermi surface are quite different.
It can be seen from Fig.~\ref{fig1:Nil}(a) that for the gsb of $^{104}$Pd,
there are four $g_{9/2}$ proton holes ($\pi 9/2^+[404]$ and $\pi 7/2^+[413]$).
At rotational frequency $\hbar\omega = 0$~MeV,
they should be partly occupied due to the pairing correlations,
and their occupation probabilities will change with increasing rotational frequency.
Meanwhile, it can also be seen from Fig.~\ref{fig1:Nil}(b) that
the neutron $h_{11/2}$ orbital $\nu 1/2^-[550]$ should be nearly empty at
rotational frequency $\hbar\omega = 0$~MeV and with rotational frequency increasing,
the single particle energy of this orbital decreases quickly.
Different from the AMR band in
$^{101}$Pd~\cite{Sugawara2012_PRC86-034326,Sugawara2015_PRC92-024309,Singh2017_JPG44-075105},
the $\nu 1/2^-[550]$ orbital is not blocked in the gsb of $^{104}$Pd,
so neutron level crossing will happen with increasing rotational frequency.
The data show that the possible AMR band in $^{104}$Pd is the yrast band
after the backbending~\cite{Rather2014_PRC89-061303R}.
Using the semiclassical particle rotor model, its configuration was assumed to be
$\pi g_{9/2}^{-2} \otimes \nu[h_{11/2}^2,(g_{7/2},d_{5/2})^2]$~\cite{Rather2014_PRC89-061303R}.
However, this band was assigned a different configuration as
$\pi g_{9/2}^{-4} \otimes \nu[h_{11/2}^2,(g_{7/2},d_{5/2})^6]$
in a recent investigation using TAC-CDFT~\cite{Jia2018_PRC97-024335}.
Therefore, to clarify its configuration, in the following investigation,
adiabatic calculations for the gsb in $^{104}$Pd
will be performed and the configuration for this AMR band
can be obtained automatically after neutron level crossings.

\begin{center}
\centering
\includegraphics[width=0.9\columnwidth]{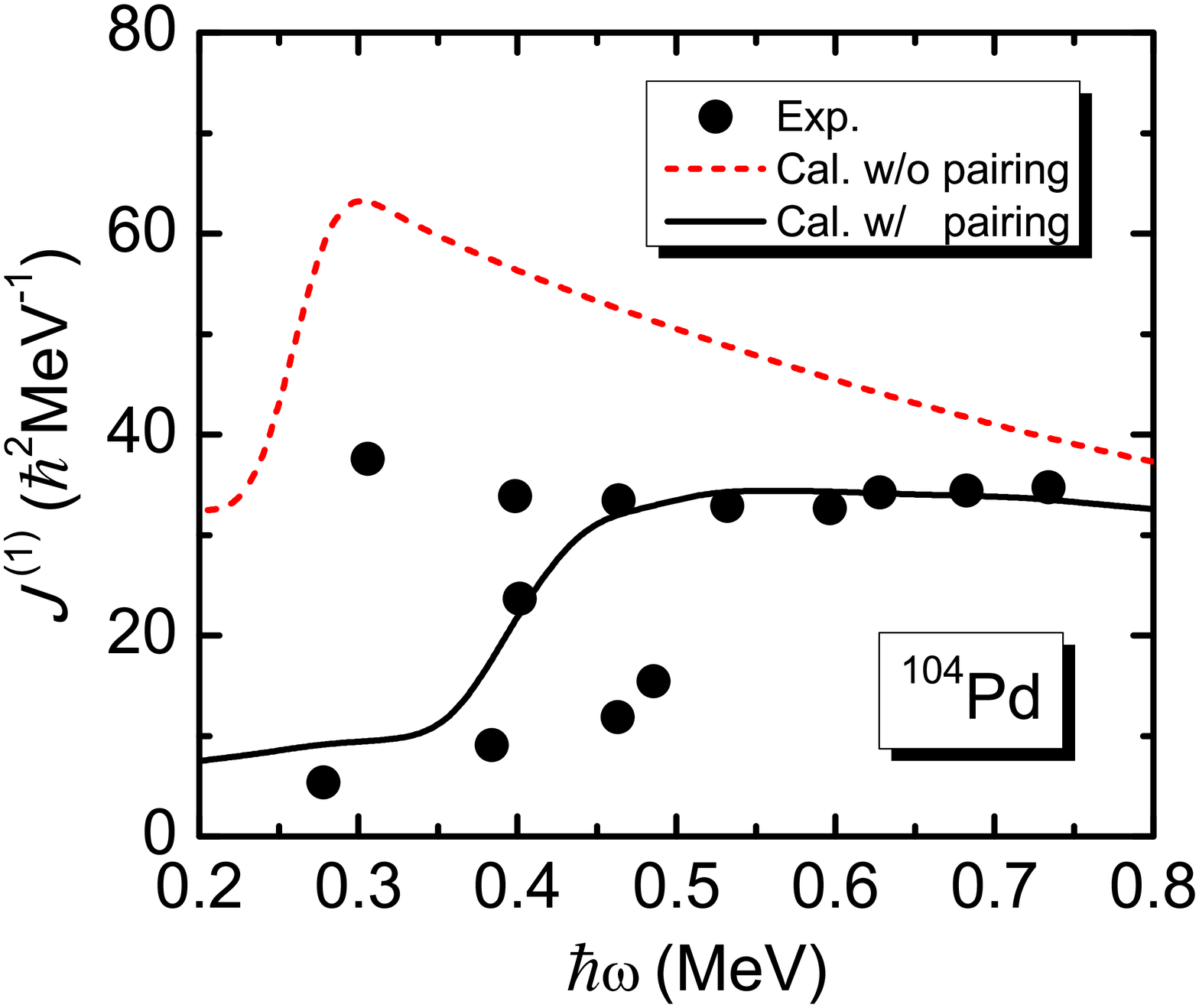}
\figcaption{\label{fig2:MOI} (Color online)
The experimental~\cite{Rather2014_PRC89-061303R} and calculated kinematic MOIs
$J^{(1)}$ with and without pairing correlations for the gsb in $^{104}$Pd.}
\end{center}

Figure~\ref{fig2:MOI} shows the experimental~\cite{Rather2014_PRC89-061303R}
and calculated kinematic MOIs $J^{(1)}$ with and without
pairing correlations for the gsb in $^{104}$Pd.
The MOIs and the corresponding rotational frequencies are extracted by
\begin{eqnarray}
\frac{J^{(1)}(I)}{\hbar^2}&=&\frac{2I+1}{E_{\gamma}(I+1\rightarrow
I-1)} \ ,  \nonumber \\ 
\hbar\omega(I)&=&\frac{E_{\gamma}(I+1\rightarrow
I-1)}{I_{x}(I+1)-I_{x}(I-1)} \ ,
\end{eqnarray}
where $I_{x}(I)=\sqrt{(I+1/2)^{2}-K^{2}}$.
The pairing correlations are crucial to reproduce
the data for the whole rotational frequency region.
One can see from Fig.~\ref{fig2:MOI} that without pairing correlations,
the MOIs of $^{104}$Pd are overestimated
for a large extent and the calculated backbending frequency is much earlier
than the data ($\hbar\omega \sim 0.4$~MeV).
After considering the pairing correlations, the data can be reproduced very well.
Note that the present calculations fail to reproduce the
sharp backbending appeared in the experimental MOIs due to the
defect of the cranking model.
In order to obtain the backbending effect exactly,
one has to go beyond the cranking model~\cite{Hamamoto1976_NPA271-15,Cwiok1978_PLB76-263}.
Due to the fact that the MOIs of the gsb in $^{104}$Pd are reproduced very well
by the present PNC-CSM calculation, the configuration of this AMR band
should be reasonable. Later on, the configuration obtained by the present calculation
will be compared with previous works~\cite{Rather2014_PRC89-061303R,Jia2018_PRC97-024335}.

Usually, typical AMR has weak $E2$ transitions and large $J^{(2)}/B(E2)$ ratios,
which reflect the nearly spherical or weakly deformed core.
Moreover, due to the ``two-shears-like'' mechanism,
the $B(E2)$ values decrease with increasing spin.
Figure~\ref{fig3:be2}(a) shows the experimental~\cite{Rather2014_PRC89-061303R} and
calculated $B(E2)$ values with and without pairing correlations for the gsb in $^{104}$Pd.
Previous works have shown that pairing correlations
are important to reproduce the $B(E2)$ values only when
the proton level crossing happens, and the reduced $B(E2)$ values strongly depend on the
deformation rather than the superfluidity~\cite{Zhang2013_PRC87-054314,Zhang2016_PRC94-034305}.
It can be seen from Fig.~\ref{fig3:be2} that the decreasing of the $B(E2)$ values
with increasing rotational frequency can be obtained no matter whether the
pairing correlations are taken into account or not.
However, the calculated results are more consistent with the data
after considering the pairing correlations.
Note that the experimental $B(E2)$ values drop very quickly with
increasing rotational frequency when $\hbar\omega>0.6$~MeV and
the present calculation with fixed deformation fails to reproduce this.
Therefore, the ``two-shears-like'' mechanism alone is not enough to provide
the decrease of $B(E2)$ values.
The PNC-CSM calculations with deformation $\varepsilon_2$ changing from 0.18 to zero
are also shown in Fig.~\ref{fig3:be2}(a) by the blue dash dotted line.
The inset shows the deformation parameter $\varepsilon_2$
adopted in this calculation, which is obtained by fitting the
experimental $B(E2)$ values with increasing rotational frequency.
It can be seen that at lower rotational frequency region, the deformation
keeps nearly unchanged. With rotational frequency increasing,
the deformation decreases gradually. When the rotational
frequency $\hbar\omega>0.65$~MeV, the deformation shows a sharp reduction
and decrease to zero rapidly at $\hbar\omega \sim 0.75$~MeV.
Due to the fact that with the experimental deformation parameter $\varepsilon_2=0.18$,
the $B(E2)$ values can be reproduced quite well at the beginning of this AMR band,
the involution of the deformation with rotational frequency should be reasonable.
Therefore, the present calculations also imply the quick reduction of the deformation
in the gsb of $^{104}$Pd with increasing rotational freuqency.
Note that the TAC-CDFT calculations in Ref.~\cite{Jia2018_PRC97-024335},
which can treat the deformation self-consistently with rotational frequency, can not
reproduce such quickly decreasing $B(E2)$ values neither.
This may need further investigation.
Figure~\ref{fig3:be2}(b) shows the comparison of
experimental and calculated $J^{(2)}/B(E2)$ ratios.
Because the $B(E2)$ values at higher spin region are not reproduced well
by the PNC-CSM calculation with fixed deformation,
only the first two data of $J^{(2)}/B(E2)$ ratios can be reproduced reasonable well.
It also can be seen that if the deformation changing effects are taken into account,
the $J^{(2)}/B(E2)$ ratios can be well reproduced,
which is shown by the blue dash dotted line in Fig.~\ref{fig3:be2}(b).
In addition, The sharp peak at $\hbar\omega \sim 0.45$~MeV for
$J^{(2)}/B(E2)$ ratios is due to the backbending.

\begin{center}
\centering
\includegraphics[width=0.9\columnwidth]{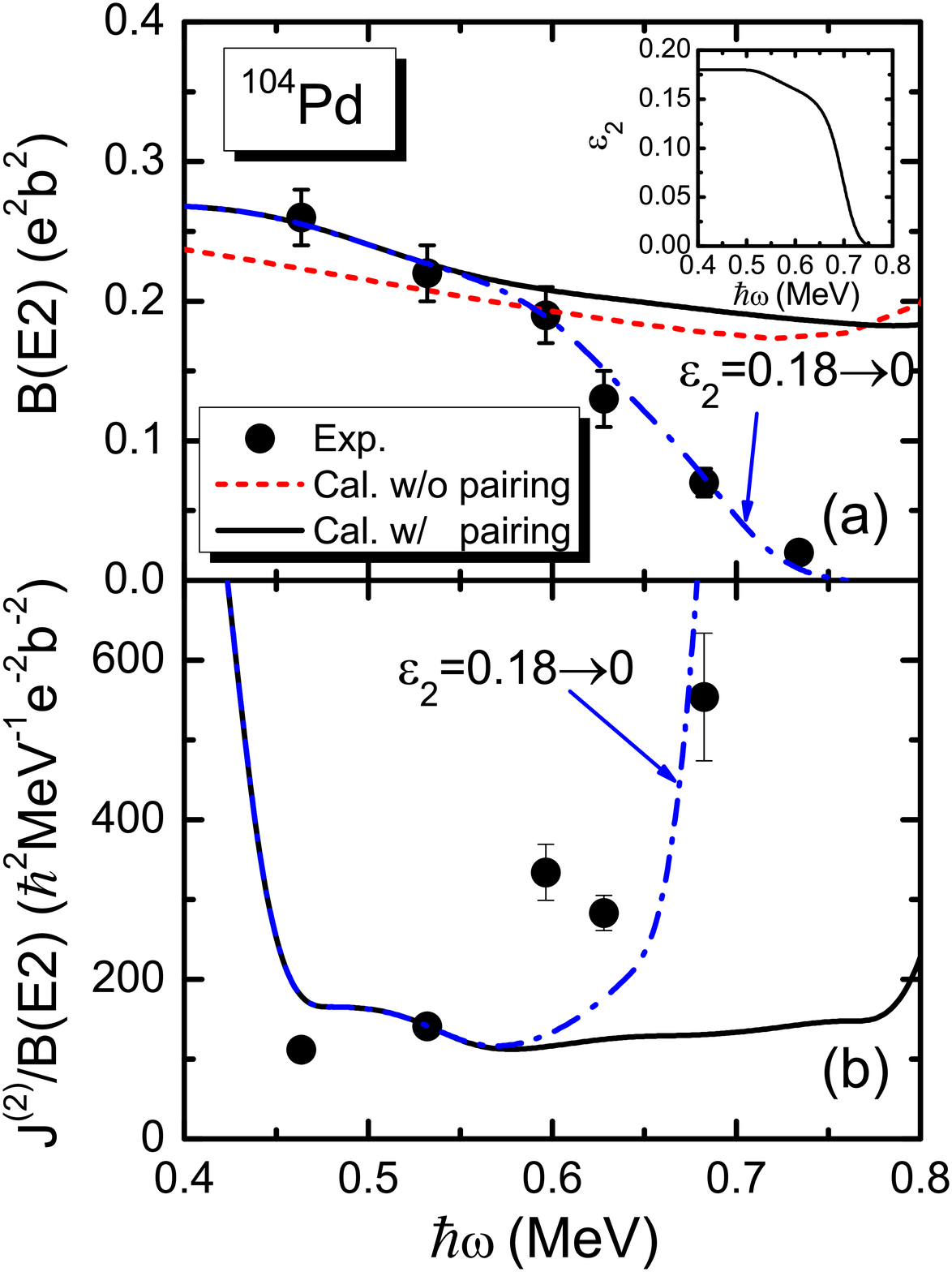}
\figcaption{\label{fig3:be2}(Color online)
(a) The experimental~\cite{Rather2014_PRC89-061303R} and calculated $B(E2)$ values
with and without pairing correlations for the gsb in $^{104}$Pd.
(b) The comparison of experimental and calculated $J^{(2)}/B(E2)$ ratios.
The blue dash dotted line shows the calculated results with
deformation $\varepsilon_2$ changing from 0.18 to zero and
the inset shows the deformation parameter $\varepsilon_2$
adopted in this calculation.}
\end{center}

Figure~\ref{fig4:occup} shows the neutron and proton occupation probability $n_\mu$ of each orbital
$\mu$ (including both $\alpha=\pm1/2$) near the Fermi surface of the gsb in $^{104}$Pd.
It can be easily seen from Fig.~\ref{fig4:occup}(a) that at
the rotational frequency $\hbar\omega \sim 0.4$ MeV,
the occupation probability for the neutron $h_{11/2}$ orbital $\nu1/2^-[550]$ increases
suddenly from about 0.4 to 2.0, while the occupation probabilities for some $(g_{7/2},d_{5/2})$
orbitals ($\nu 3/2^+[411]$, $\nu 5/2^+[413]$, etc.) decrease.
This indicates that the backbending may mainly be caused by
the alignment of one pair of $h_{11/2}$ neutron.
Therefore, the AMR, in which the increase of angular momentum
is caused by the contribution from proton,
can happen in $^{104}$Pd only after the neutron level crossing.
It also can be seen from Fig.~\ref{fig4:occup}(b) that at the similar
rotational frequency, the occupation probability for
the proton $g_{9/2}$ orbital $\pi 7/2^+[413]$ drops down
from about 0.6 to about 0.1, while the occupation probabilities for other proton $g_{9/2}$ orbitals,
e.g., $\pi 5/2^+[422]$ and $\pi 3/2^+[431]$, increase gradually.
The occupation rearrangements in proton $g_{9/2}$ orbitals may also contribute
to the backbending, which is similar as $^{101}$Pd~\cite{Zhang2016_PRC94-034305}.
The present calculations show that after backbending,
due to the existing of pairing correlations,
the proton configuration of the AMR band is four
partly empty proton $g_{9/2}$ holes.
As for the neutron, the configuration is two aligned $h_{11/2}$
particles and about six particles in the $(g_{7/2},d_{5/2})$ orbitals.
Note that the proton holes and neutron particles are written
with respect to the $^{100}$Sn ($N=50$ and $Z=50$) core.
When neglecting the pairing correlations,
this configuration is consistent with
that adopted in Ref.~\cite{Jia2018_PRC97-024335} by the TAC-CDFT.
In addition, one can see from Fig.~\ref{fig4:occup}(b) that at
the rotational frequency $\hbar\omega \sim 0.8$ MeV,
the occupation of $\pi 1/2^+[431]$ increases from nearly zero to
about 0.6, and the occupation of $\pi 5/2^+[422]$ drops from 1.9 to
1.3. This is caused by the level crossing between these two orbitals due
to the quick drop of $\pi 1/2^+[431]$ with increasing rotational frequency,
which can be  easily seen from the cranked Nilsson levels in
Fig.~\ref{fig1:Nil}(a).
This indicates that the AMR band observed in $^{104}$Pd will be terminated
around this rotational frequency and a four quasiparticle band will appear,
which is similar as that observed in $^{143}$Eu~\cite{Rajbanshi2015_PLB748-387}.

\begin{center}
\centering
\includegraphics[width=0.9\columnwidth]{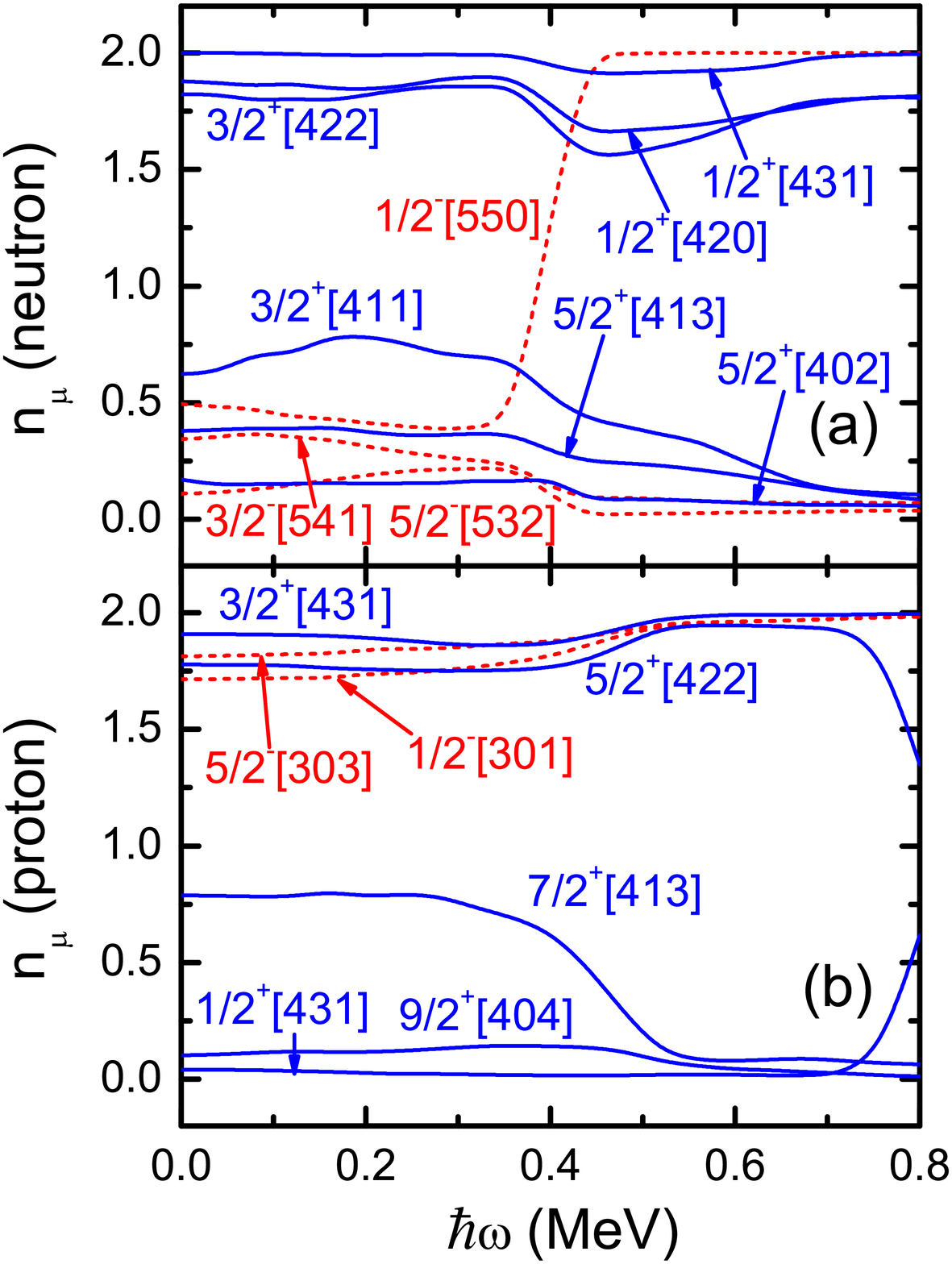}
\figcaption{\label{fig4:occup}(Color online)
The neutron (a) and proton (b) occupation probability $n_\mu$ of each orbital
$\mu$ (including both $\alpha=\pm1/2$) close to the Fermi surface
of the gsb in $^{104}$Pd.
The positive and negative parity levels are denoted by blue solid and red dashed lines.
The Nilsson levels which are fully occupied
($n_{\mu}\sim0$) and fully empty ($n_{\mu}\sim2$) are not shown.}
\end{center}

Figure~\ref{fig5:jx} shows
the experimental and calculated
angular momentum alignment $\langle J_x\rangle$ for the gsb in $^{104}$Pd.
It can be seen from Fig.~\ref{fig5:jx} that,
there is a sharp increase of the neutron angular momentum alignment
from rotational frequency $\hbar\omega \sim 0.35$~MeV to about 0.45~MeV,
which indicates again that this backbending mainly comes from the contribution of neutrons.
Meanwhile, from $\hbar\omega \sim 0.40$~MeV to about 0.55~MeV,
the contribution from protons to $\langle J_x\rangle$ also increases gradually.
This increase comes from the rearrangement of proton occupations in $g_{9/2}$ orbitals,
which is similar as that in $^{101}$Pd~\cite{Zhang2016_PRC94-034305}.
It can be seen that before finishing the rearrangement of proton occupations
in $g_{9/2}$ orbitals between $\hbar\omega \sim 0.45$~MeV to about 0.55~MeV,
the contribution from the proton to $\langle J_x\rangle$ is larger than the neutron.
This indicates that the contribution from AMR to the increasing of $\langle J_x\rangle$
is larger than that from the normal rotation.
At $\hbar\omega > 0.55$~MeV, the increase of neutron angular
momentum alignment is faster than that of proton.
This indicates that the contribution from AMR to the increasing of $\langle J_x\rangle$
is smaller than that from the normal rotation.
Therefore, the AMR in $^{104}$Pd may separate into two stages.

\begin{center}
\centering
\includegraphics[width=0.9\columnwidth]{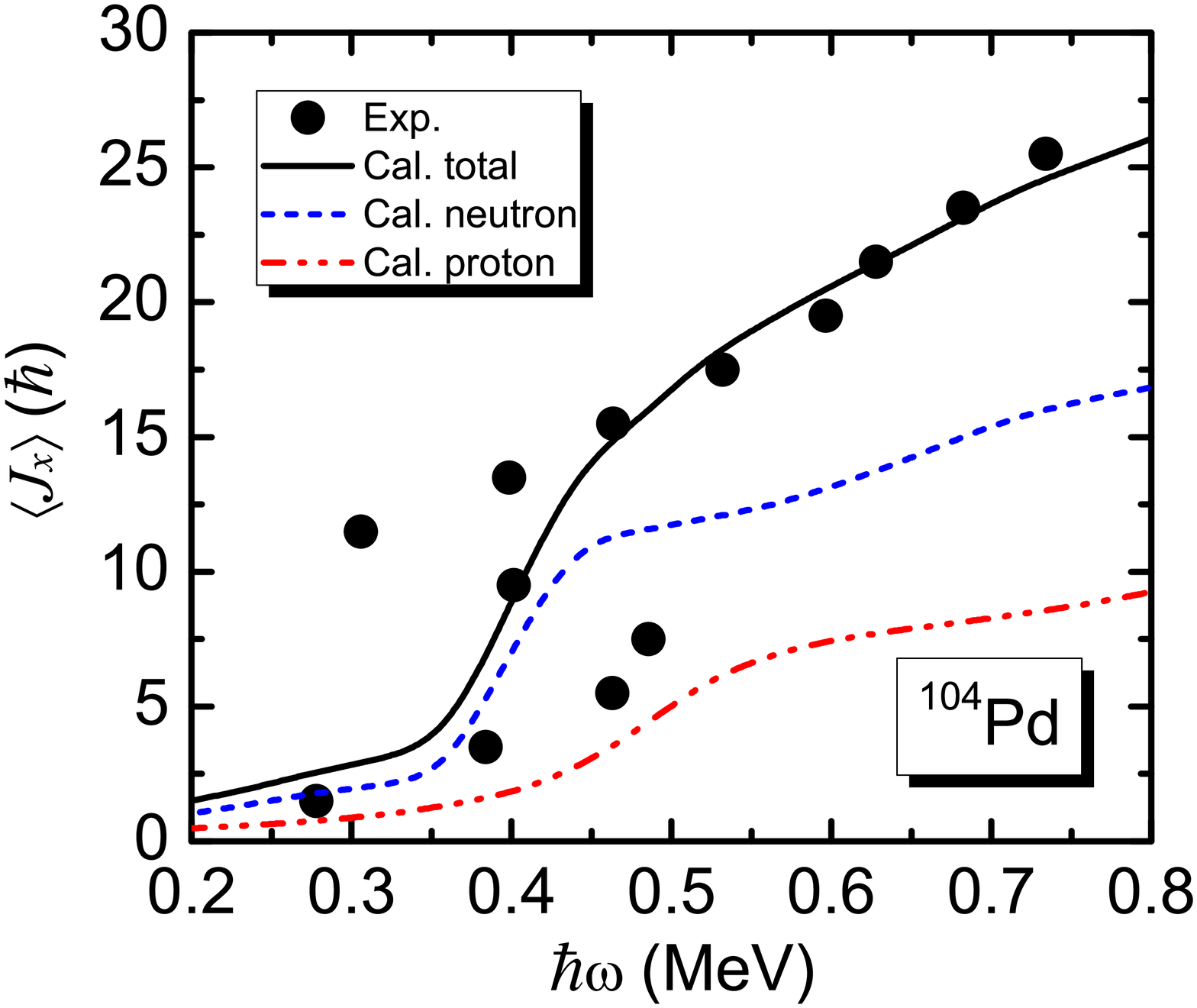}
\figcaption{\label{fig5:jx}(Color online)
The experimental and calculated
angular momentum alignment $\langle J_x\rangle$ for the gsb in $^{104}$Pd.}
\end{center}

To see the  neutron and proton alignment process more clearly,
the contributions of neutron and proton $N=4$ and 5 major shells
to the angular momentum alignment $\langle J_x\rangle$ for the gsb
in $^{104}$Pd are shown in Fig.~\ref{fig6:jxshell}.
The contributions of diagonal $\sum_{\mu} j_x(\mu)$ and off-diagonal part
$\sum_{\mu<\nu} j_x(\mu\nu)$ in Eq.~(\protect\ref{eq:j1})
from the neutron $N=5$ and proton $N=4$ major shell are denoted by dashed lines.
It can be seen that the backbending around $\hbar\omega\sim$ 0.4~MeV
mainly comes from the neutron $N=5$ major shell, especially from the diagonal part.
Furthermore, if one looks into the details,
the neutron diagonal part $j_x\left(\nu 1/2^-[550]\right)$
is mainly responsible for the backbending.
In addition, the neutron $N=4$ major shell provides a gradual increase of the
angular momentum alignments after the neutron level crossing.
This originates from the contribution of those neutron $(g_{7/2},d_{5/2})$ orbitals close to the Fermi surface.
Fig.~\ref{fig6:jxshell}(b) shows that the proton $N=4$ major shell also contributes a gradual increase of the
angular momentum alignment around the backbending frequency $\hbar\omega\sim$ 0.4~MeV,
in which the off-diagonal part contributes a lot.
Furthermore, the proton off-diagonal parts
$j_x\left(\pi 5/2^+[422] \pi 7/2^+[413]\right)$, and
$j_x\left(\pi 7/2^+[413] \pi 9/2^+[404]\right)$
are mainly responsible for this gradual increase in proton angular momentum alignment.
It also can be seen that at higher rotational frequency region with $\hbar\omega > 0.55$~MeV,
the increase of the angular momentum alignment from proton $N=4$ major shell is
slower than that from the neutron $N=4$ major shell.
This again tell us that the increase of the angular momentum alignment from AMR is less than
that provided by the neutron $(g_{7/2},d_{5/2})$ orbitals at higher rotational frequency region.

\begin{center}
\centering
\includegraphics[width=0.9\columnwidth]{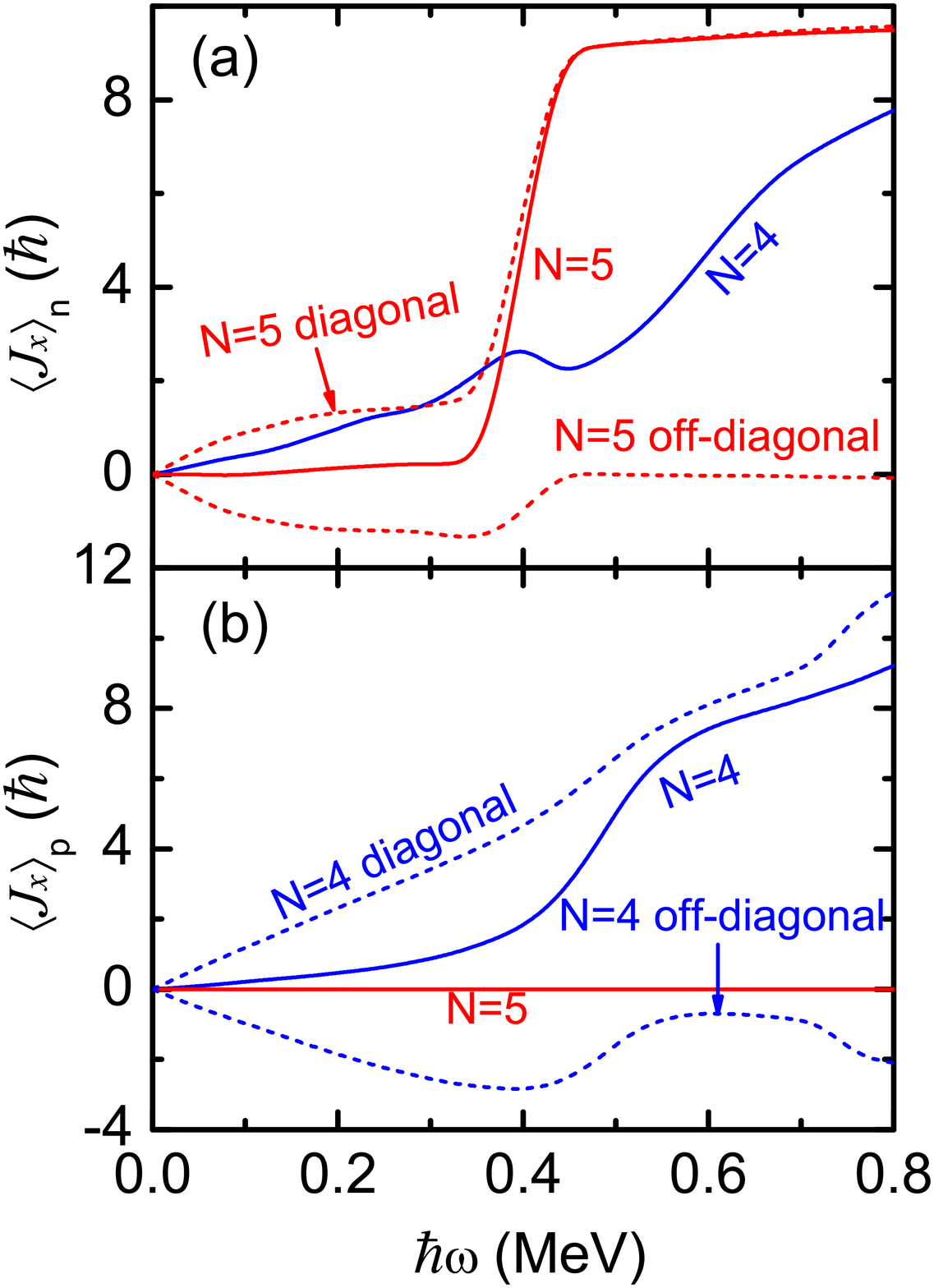}
\figcaption{\label{fig6:jxshell}(Color online)
Contributions of (a) neutrons and (b) protons $N=4$ and 5 major shells
to the angular momentum alignment $\langle J_x\rangle$ for
the gsb in $^{104}$Pd.
Contributions of diagonal $\sum_{\mu} j_x(\mu)$ and off-diagonal part
$\sum_{\mu<\nu} j_x(\mu\nu)$ (c.f., Eq.~\ref{eq:j1})
from the neutron $N=5$ and proton $N=4$ major shells are denoted by dashed lines.}
\end{center}

\begin{center}
\centering
\includegraphics[width=0.9\columnwidth]{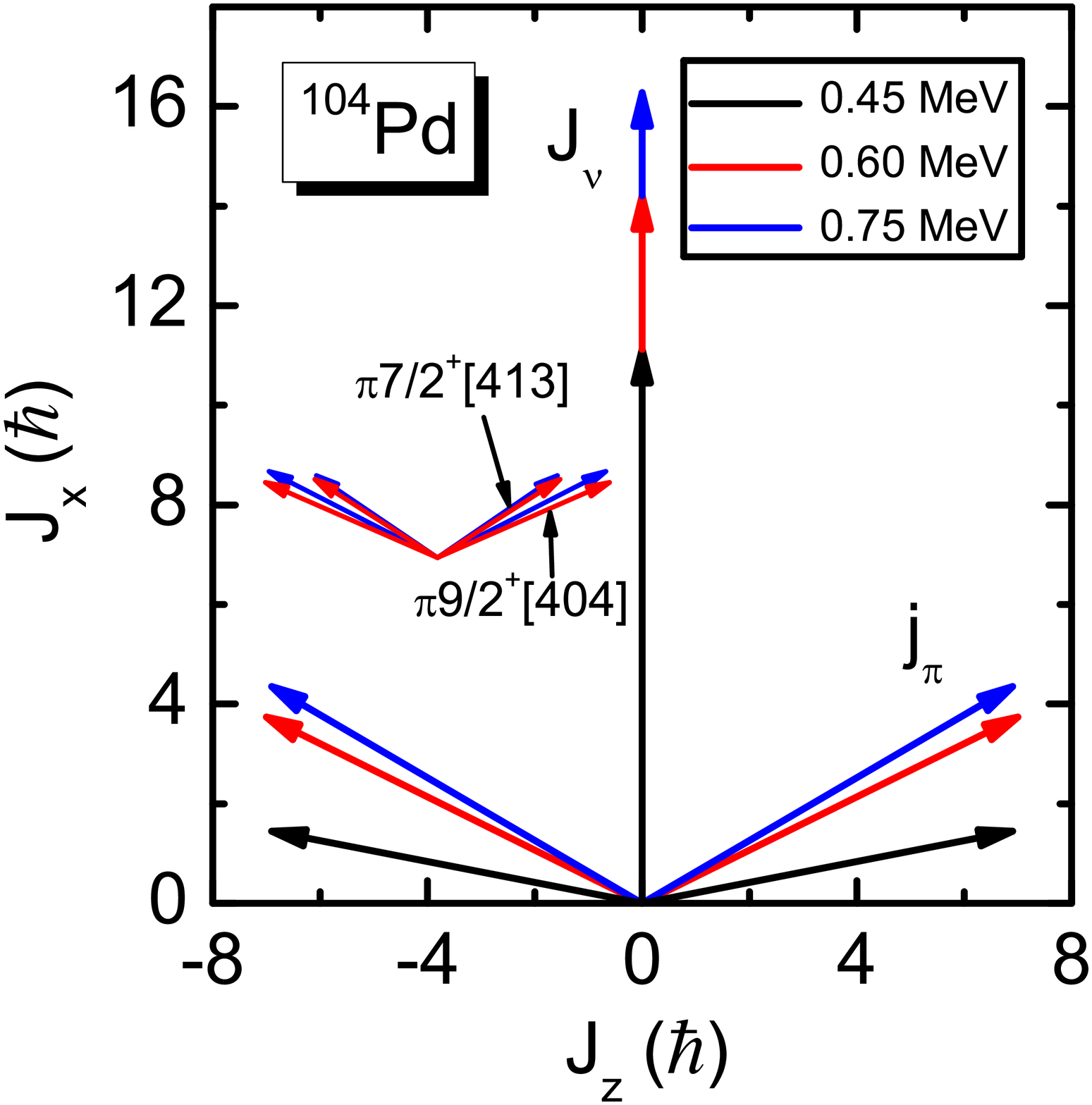}
\figcaption{\label{fig7:shear}(Color online)
Angular momentum vectors of  four proton $g_{9/2}$ holes ($j_\pi$)
and neutrons ($J_\nu$) for the gsb in $^{104}$Pd.
Each $j_\pi$ includes the contribution from two proton $g_{9/2}$ holes.
The inset shows the angular momentum vectors of
four proton $g_{9/2}$ holes ($\pi9/2^+[404]$ and $\pi7/2^+[413]$) separately
by neglecting the off-diagonal part in Eq.~\ref{eq:j1}. }
\end{center}

Figure~\ref{fig7:shear} shows the angular momentum vectors of
four proton $g_{9/2}$ holes ($j_\pi$) and neutrons ($J_\nu$) for
the gsb in $^{104}$Pd at rotational frequencies from 0.45 to 0.75~MeV.
Each $j_\pi$ includes the contribution from two proton $g_{9/2}$ holes.
Note that due to the pairing correlations,
the contribution from the off-diagonal part
$\sum_{\mu<\nu} j_x(\mu\nu)$ of protons is none zero (c.f., Fig.~\ref{fig6:jxshell}),
especially at lower rotational frequency region ($\hbar\omega$=0.45 to 0.60~MeV).
Therefore, it is difficult to separate the contribution of each proton $g_{9/2}$
hole to the angular momentum alignment to get the ``umbrella''-like
AMR mode as in Ref.~\cite{Jia2018_PRC97-024335}.
However, with rotational frequency $\hbar\omega>0.60$~MeV,
the four proton $g_{9/2}$ holes are nearly empty and the
contribution from the off-diagonal part to $J_x$ becomes smaller.
Therefore, we can separate the angular momenta of these four $g_{9/2}$
holes approximately by neglecting the contribution from the off-diagonal part to $J_x$,
which has been shown as an inset in Fig.~\ref{fig7:shear}.
According to Ref.~\cite{Frauendorf1996_ZPA356-263},
$J_z$ is calculated approximately by
\begin{equation}
J_z = \sqrt{\langle \Psi | J_z^2 | \Psi \rangle}
\end{equation}
in the present PNC-CSM formalism.
By comparing the principal axis cranking with the particle rotor model,
this method has already been demonstrated to be a good approximation.
One can see from Fig.~\ref{fig7:shear} that at $\hbar\omega = 0.45$~MeV,
proton vectors $j_\pi$ are pointing opposite direction and are nearly
perpendicular to the neutron vector $J_\nu$.
With rotational frequency increasing, the two proton $j_\pi$ vectors
gradually close toward the neutron $J_\nu$ vector,
while the direction of the total angular momentum stays unchanged.
Therefore, higher angular momentum is generated due to the ``two-shears-like'' mechanism.
From rotational frequency $\hbar\omega =0.45$ to 0.75~MeV,
the neutron angular momentum alignment increases smoothly.
This is due to the alignment of neutrons in $g_{7/2}$ and $d_{5/2}$ orbitals,
which is consistent with the TAC-CDFT calculations in Ref.~\cite{Jia2018_PRC97-024335}.
The two proton blades close rapidly with rotational frequency
$\hbar\omega =0.45$ to 0.60~MeV,
and the magnitude of two $j_\pi$ vectors keep no longer constant.
This is similar as $^{101}$Pd~\cite{Zhang2016_PRC94-034305},
which originates from the occupation rearrangement in proton $g_{9/2}$ orbitals.
It should be noted that since pairing correlations are neglected
in the TAC-CDFT calculations for $^{104}$Pd in Ref.~\cite{Jia2018_PRC97-024335},
the four proton $g_{9/2}$ holes are fully empty, and there is no
proton occupation rearrangement process with increasing rotational frequency.
Therefore, the proton blades with nearly constant magnitude close steadily with increasing rotational frequency,
which is quite different from the angular momenta picture in the present PNC-CSM calculation.
With rotational frequency increasing from 0.60 to 0.75~MeV,
the close of two proton blades become slow and steady,
which is similar with the typical AMR in $^{105, 106}$Cd~\cite{Zhang2013_PRC87-054314}.
Therefore, two stages of AMR in $^{104}$Pd are clearly seen.
It also can be seen from the inset of Fig.~\ref{fig7:shear} that with the magnitude keeping
constant, the angular momentum vectors of four proton $g_{9/2}$
holes ($\pi9/2^+[404]$ and $\pi7/2^+[413]$) close simultaneously with increasing rotational frequency.
In addition, the two $\pi9/2^+[404]$ blades
close more quickly than $\pi7/2^+[413]$ blades, which indicates that the two
$\pi9/2^+[404]$ holes contribute more angular momenta in the typical AMR mode.

\section{Summary}{\label{Sec:Summary}} \vspace*{-1mm}

In summary, the particle-number-conserving method based on the cranked shell model
is used to investigate the AMR band in $^{104}$Pd.
The experimental MOIs are reproduced quite well.
In order to reproduce the $B(E2)$ values,
a corresponding deformation change with increasing
rotational frequency is necessary.
The $J^{(2)}/B(E2)$ ratios has also been discussed.
The occupation probability of each orbital close to the Fermi surface
and the contribution of each major shell to the total angular momentum
alignment with rotational frequency are analyzed.
The backbending mechanism of the ground state band in $^{104}$Pd is understood clearly
and the configuration of the AMR after backbending is clarified.
The present calculations suggest that the configuration of the AMR band
in $^{104}$Pd should be $\pi g_{9/2}^{-4} \otimes \nu[h_{11/2}^2,(g_{7/2},d_{5/2})^6]$
when the pairing correlations are neglected,
which is consistent with the TAC-CDFT calculations with PC-PK1.
Furthermore, the crossing of a four quasiparticle states
with this AMR band is also predicted.
Finally, the ``two-shears-like'' mechanism for the AMR is
investigated by examining the shears angle, and two stages of AMR in
$^{104}$Pd are clearly seen.


\end{multicols}

\vspace{-1mm}
\centerline{\rule{80mm}{0.1pt}}
\vspace{2mm}

\begin{multicols}{2}

\end{multicols}

\clearpage
\end{CJK*}

\end{document}